# Dynamic electrophysical characterization of rubidium polytungstate ceramics under fast humidity impact

S. L. Bravina[*,1,1], N. V. Morozovsky[1], S. F. Solodovnikov[2],

O. M. Basovich[3] and E. G. Khaikina[3]

[1]*Institute of Physics of NAS of Ukraine, 46, prosp. Nauki, Kiev, 03028 Ukraine*

[2]*A.V. Nikolaev Institute of Inorganic Chemistry, Siberian Branch, Russian Academy of Sciences, 3, Acad. Lavrent'ev Ave., Novosibirsk, 630090 Russia*

[3]*Baikal Institute of Nature Management, Siberian Branch, Russian Academy of Sciences, 8, Sakh'yanova Str., Ulan-Ude, 670047 Russia*

**Abstract.** The electrophysical reaction of rubidium polytungstate $Rb_4W_{11}O_{35}$ ceramics (RPTC) on fast humidity impact has been investigated by studying the influence of wet air pulse of 1 – 3 s of duration on the peculiarities of dynamic current-voltage (I-V-) and charge-voltage (Q-V-) loops and transient current-time (I-t-) curves.

The revealed peculiarities of I-V- and Q-V-loops and I-t-curves for RPTC subjected to humidity impact were compared with similar ones observed for water damaged hydrogen-bounded ferroelectric triglicyne sulphate (TGS) crystals.

The obtained data indicate at least two possible scenarios of polar reaction on the humidity impact, namely protonic/ionic space charge transfer and ferroelectric-like reversible polarization in surface conglomerate of RPTC disturbed by dipolar $H_2O$ adsorbate.

Due to a high surface activity the RPTC can be considered as promising materials for creating humidity sensitive elements of environment monitoring systems. Direct humidity-to-frequency (time) conversion was realized by inserting metal-RPTC-metal humidity sensitive element in LC resonance or RC-relaxation frequency control circuits of generating units.

I. **Introduction**

Simple and complex crystalline tungstates are considered as functional materials for nonlinear optic (Kaminskii 1984) and scintillation (Baccaro *et al.* 1998, Juestel and Ronda 2010) devices. Crystalline tungstates possess also ferroelectric properties (Lines and Glass 1977, Isupov 2005).

---

[1] Corresponding author. E-mail: bravmorozo@yahoo.com

Ceramics of polytungstates can be considered as attractive for fundamental and application directed investigations in humidity sensing area along with other metal-oxide ceramics (Kulwicki 1991, Traversa 1995), mixed molibdate-tungstate systems (Barinova and Kirsanova 2008), titanates (Gajovića *et al.* 2009), and simple tungstates (You *et al.* 2012).

Taking into account a tendency of integrating surface-active structures into functional electronics and the demand for fast response, the study of impact of humidity changes on operation of surface-active materials not only in static but also dynamic modes should be a subject of special interest.

Earlier (Bravina *et al.* 1999, Bravina *et al.* 2000, Bravina and Morozovsky 2002, Bravina *et. al.* 2006, Bravina *et al.* 2009) we revealed the significant and fast humidity sensitivity of a number of porous systems. Together with zeolite-like Na-Y and mesoporous MCM-41 systems (Bravina *et al.* 1999) and porous Si (Bravina *et al.* 2006, Bravina *et al.* 2009) some complex metal-oxide ceramics were reported as humidity sensitive materials (Bravina *et al.* 2000, Bravina *et al.* 2009). These studies proved the efficiency of examining the changes of parameters of bipolar and unipolar dynamic current-voltage and charge-voltage characteristics and transient currents connected with the pulse change of humidity.

In this paper we present the results of examining pulse humidity impact on dynamic electrophysical characteristics of rubidium polytungstate, $Rb_4W_{11}O_{35}$, ceramics (RPTC).

## II. Experimental
### II.1. Samples

The complex oxide compounds $Rb_4W_{11}O_{35}$ in polycrystalline state were synthesized by solid state reactions. Polytungstate of rubidium, $Rb_4W_{11}O_{35}$, was synthesized according to the reaction $2\ Rb_2CO_3 + 11\ WO_3 = Rb_4W_{11}O_{35} + 2\ CO_2$. The reaction mixture was stepwise annealed in air at 600–850°C during 80 h.

Single phase state of a synthesized compound was confirmed by powder X-ray diffraction on automated D8 ADVANCE Bruker diffractometer ($CuK_\alpha$ radiation, secondary monochromator, maximal angle $2\theta = 100°$, scan step $0.02°$). Primary X-ray data processing was performed using PROFAN CSD software package.

$Rb_4W_{11}O_{35}$ undergoes incongruent melting at 990°C, crystallizes in the orthorhombic system ($a = 14.635(1)$, $b = 25.659(1)$, $c = 7.675(1)$ Å, possible space groups $Pca2_1$, $Pnc2_1$, Z

= 4) as a superstructure relative to the hexagonal tungsten bronze type.

The projection of $Rb_4W_{11}O_{35}$ structure on (010) plane of monoclinic sub-cell is presented in Figure 1. The superstructure is formed by ordered extraction of the part of $WO_6$-octahedra from W-O-framework of prototype and a crystallographic shear in the extraction plane (Solodovnikov *et.al* 1998, Solodovnikov *et al.* 2002). The tunnels formed by WO6-octahedra and interlayer spaces are occupied by Rb+ ions, at that the positions of interlayer Rb(3) are half-occupied (the distances Rb(3)-Rb(3) have usual values).

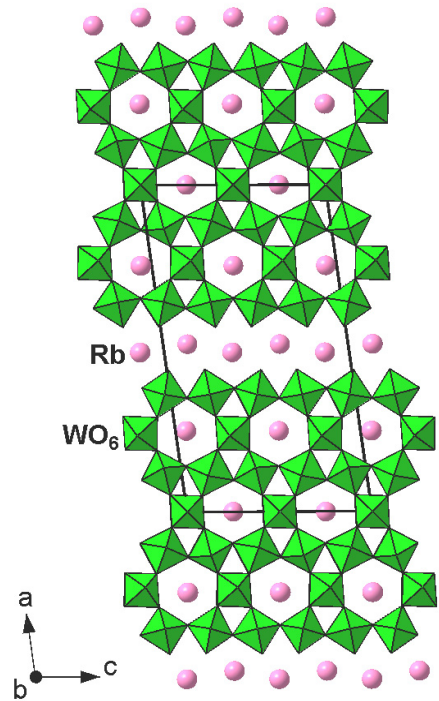

**Figure 1.** The $Rb_4W_{11}O_{35}$ structure projection on (010) plane of monoclinic sub-cell (shown by solid line). $WO_6$-octahedrons are marked by tetrahedrons (in green), Rb atoms are shown by balls (in lilac).

The RPTC were sintered from the synthesized polycrystalline compound. The samples of ≈ 0.5 mm thickness and 0.2-0.5 $cm^2$ of area were electroded by Ag-paste and delicately clamped between Cu wires.

As an object of comparison relatively to possible ferroelectric activity were selected the single crystalline plates of hygroscopic hydrogen bounded ferroelectrics triglicyne sulphate $(NH_2CH_2COOH)_3·H_2SO_4$ (TGS).

**II.2. Measurements**

The wet air flux pulses of 1-5 sec of duration and relative humidity $H_r$ near of 100 % were directed on the investigated samples.

For examining effect of step-like change of humidity on the electrical parameters of investigated systems the oscilloscopic observations of humidity impact induced variations of parameters of dynamic current-voltage (I-V-) loops, charge-voltage (Q-V-) loops and also transient current-time (I-t-) curves were performed**.**

I-V- and Q-V- loops were registered under triangular voltage and I-t-curves were registered under rectangular meander voltage as it is performed for ferroelectric (Lines and Glass 1977, Burfoot 1967, Burfoot and Taylor 1979) and metal - silicon oxide - metal (Chou 1971, Kuhn 1971) structures.

The I-V-, Q-V- and I-t- measurements were performed in the multi-cycle mode under applied drive voltage in the amplitude range $0.1 \leq V_d \leq 10$ V and frequency range 1 Hz $\leq f_d \leq$ 1 kHz.

The temporal variations of I-V- and Q-V- loops and I-t-curves were examined during and just after wet air pulse and under restoration of the initial state.

### III. Results and comments
**III.1. Influence of drive voltage amplitude and frequency**

Figure 2 presents the bipolar I-V-loops obtained for $Rb_4W_{11}O_{35}$ samples under different amplitudes and frequencies of applied drive voltage at relatively low ($H_r = 50$ %) humidity.

At sound frequencies ($f_d = 200$ Hz) the shape of low-voltage I-V-loops (Fig. 2, left) corresponds to a linear equivalent series - parallel RC-circuit (Vollman and Waser 1991, Bravina *et al.* 1991). At that, drive voltage increase results in R- and C- voltage non-linearities. Under frequency decreasing R- and C- voltage non-linearities start to be more pronounced. At infra-low frequencies ($f_d = 2$ Hz) I-V-loops (Fig. 2, right) display high R- and C- non-linearity with the "hump" regions observed under drive voltage increasing run.

On the obtained unipolar I-V-curves (Figure 3) three characteristic regions can be distinguished. The first one at low $V_d$ values is sharp, jump-like and expands with $f_d$ increase that is connected with displacement current observed at the moment of $V_d$ sew-tooth pulse termination (see Fig. 3, bottom). The second region at intermediate $V_d$ values is sub-linear and quasi-saturated. The third one at high $V_d$ values is super-linear. Under $f_d$ change from 2 to 200 Hz (see Fig. 3, right), the increase of current value and slope of I-V-curve is observed but the main details of the shape are conserved.

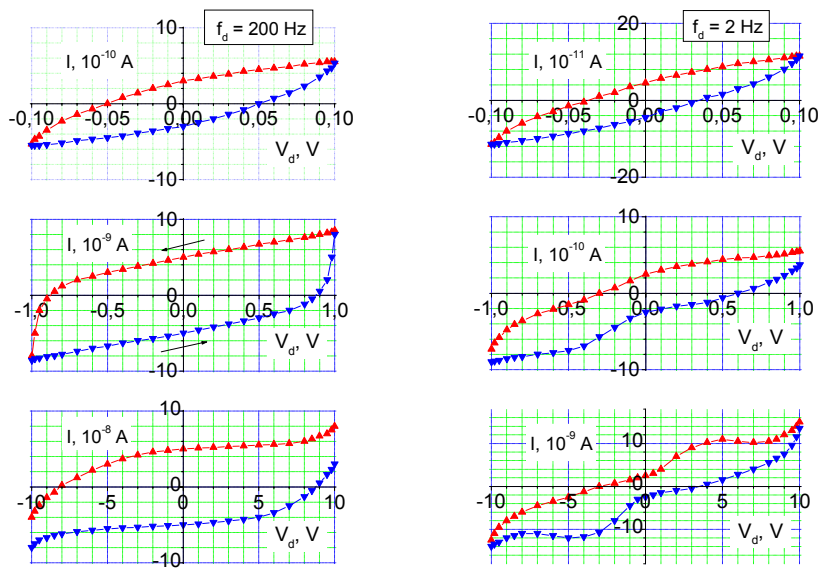

**Figure 2.** Bipolar current-voltage loops under drive voltage amplitudes 0.1, 1 and 10 V at 200 Hz **(left)** and at 2 Hz **(right).**

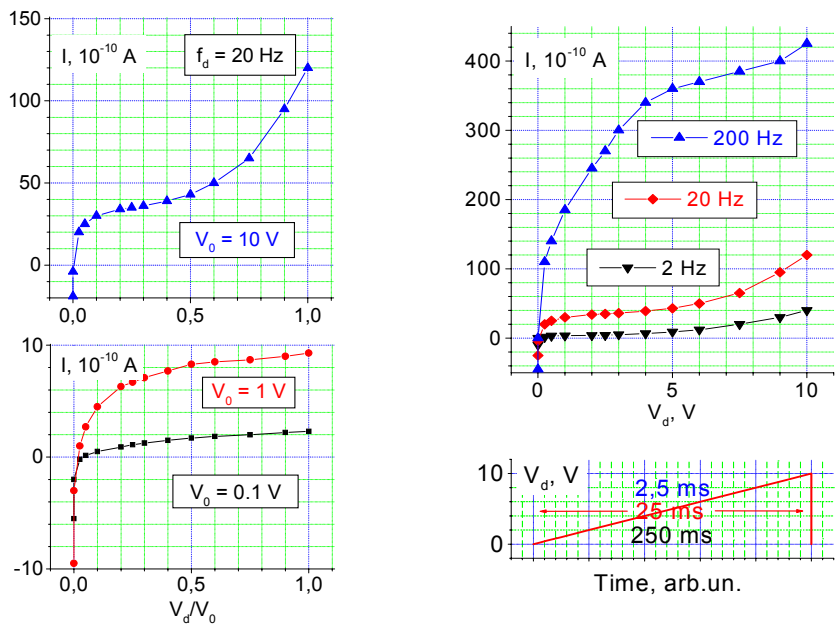

**Figure 3.** Unipolar current-voltage characterstics:
 **(left)** at 20 Hz under drive voltages amplitudes 0.1, 1 and 10 V;
 **(right)** at 10 V under drive voltage frequencies 200, 20, and 2 Hz (top) and schematic representation of sew-tooth like pulse of drive voltage (bottom).

On the I-t-curves (Figure 4) under increasing drive voltage amplitude the maxima appearance is observed that is similar to "humps" occurrence on the I-V-loops (see Fig. 2, right).

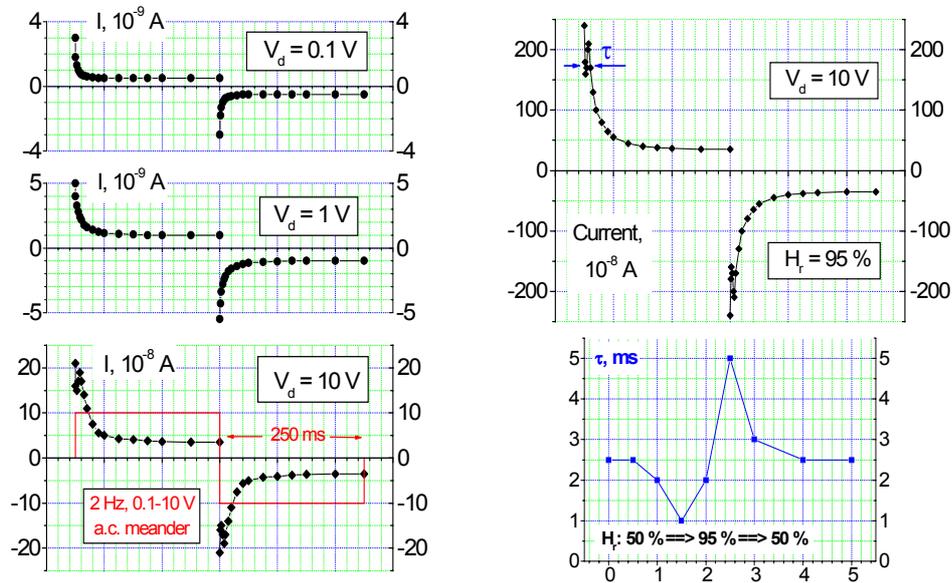

**Figure 4.** Bipolar transient currents for different drive voltage amplitudes:
 (**left**) at relative humidity 50 % and with schematic representation of meander drive voltage (bottom);
 (**right**) just after wet air pulse (relative humidity 95 %) for 10 V drive voltage amplitude and the change of maxima half-width during wet air pulse (bottom).

### III.2. Wet air pulse impact

Figure 4 presents I-t-curves and Figure 5 presents I-V-loops obtained before and just after wet air pulse impact on $Rb_4W_{11}O_{35}$ samples.

Under wet air pulse action we observed the increase of I-t-curves swing and relaxation time followed by changing the maxima width (see Fig. 4, right) and also the increase of I-V-loops average slope and width followed by increase of the "humps" height (see Fig. 5). Such transformations of I-t-curves and I-V-loops correspond to decrease of R value and increase of C value under occurrence of apparent R(V)- and C(V)- non-linearities. At that, the response time $\tau_1$ on a wet air pulse is $\tau_1 \sim 1$ s.

During and just after the wet air pulse non-monotonic change of the maxima width on I-t-curve was revealed (see Fig. 4, right).

Typical transformation of Q-V-loops obtained under wet air pulse impact is presented in Figure 5. The increase of humidity leads to the increase of the loop size along Q-axis. This corresponds to the transformation of I-V-loops and I-t-curves and reflects the increase of the value of transferred electrical charge.

The state with high C and low R values remains during the "retention" time $\tau_2 \sim 1 - 10$ s depending on wet air pulse duration. The height of the "humps" is maximal at high $H_r$ and is decreasing under drying of RPTC sample in the course of restoration of the initial state. For I-V-loops this restoration is also accompanied by decrease of degree of R- and C- non-linearity to the initial value.

After wet air pulse termination and restoration of initial low $H_r$ value the shape of I-V- and Q-V- loops and I-t-curves and their parameters are being restored with time to their initial ones. At that, the value of recovery time $\tau_3$ of restoring the initial state is $\tau_3 \sim 100$ s.

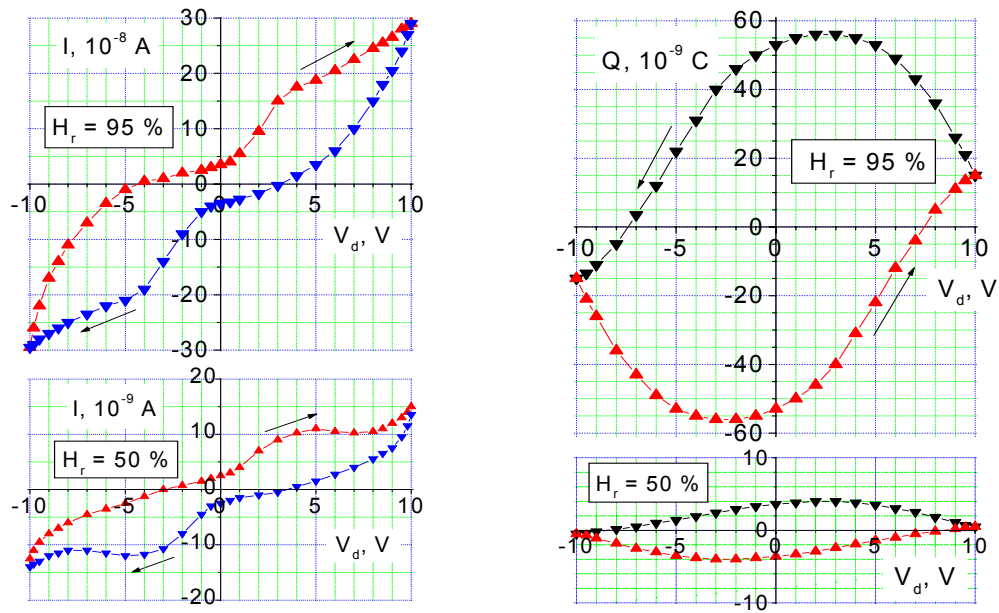

**Figure 5.** Bipolar current-voltage loops (**left**) and charge-voltage loops (**right**) before (relative humidity 50 %) and just after wet air pulse (relative humidity 95 %)**.**

## IV. Discussion
### IV.1. I-V-loops peculiarities

The main observed peculiarities of I-V-loops can be explained by simplified sample presentation with parallel R‖C equivalent circuit. For the selected parallel RC–circuit

$$I(V) = I_R(V) + I_C(V) = V/R + \partial(CV)/\partial t.$$

Under linear $V(t)$ sweeping $V(t) = V_0 - bt$ or $V(t) = -V_0 + bt$ with $dV/dt = \pm b = \text{const}$. The assumption $R = \text{const}$, $C = \text{const}$ gives $I(V)$ of parallelogram-like shape which size along V-axis is $2V_0$, the size along I-axis is proportional to C value and the slope is proportional to $1/R$ value. So, any deviation of $I(V)$ from linearity is connected with existence of some $R(V)$ and/or $C(V)$ dependences.

The "hump"-like peculiarities of I-V-loops of polarization reversal and I-t-curves of polarization switching for ferroelectric materials are connected with displacement current maxima $I_D = (\partial P/\partial V)(\partial V/dt)$ and $I_D = \partial P/\partial t$ respectively (Burfoot 1967, Burfoot and Taylor 1979). Hysteretic voltage dependence of polarization $P(V)$ under its reversal by external voltage $V(t) \sim \pm bt$ corresponds to two opposite maxima of $\partial P/\partial V$ and so $I_D = \pm b(\partial P/\partial V)$.

Summing up $I_R(V) + I_C(V)$ and $I_D(V)$ and taking into account $I_R(V)$ and $I_C(V)$ nonlinearities due to $R(V)$ and/or $C(V)$ dependences we obtain the transition from distorted parallelogram-like loop observed for low $V_d$ values (Fig. 2) to I-V- loop with two opposite maxima observed for high $V_d$ values (Figs 2 and 5).

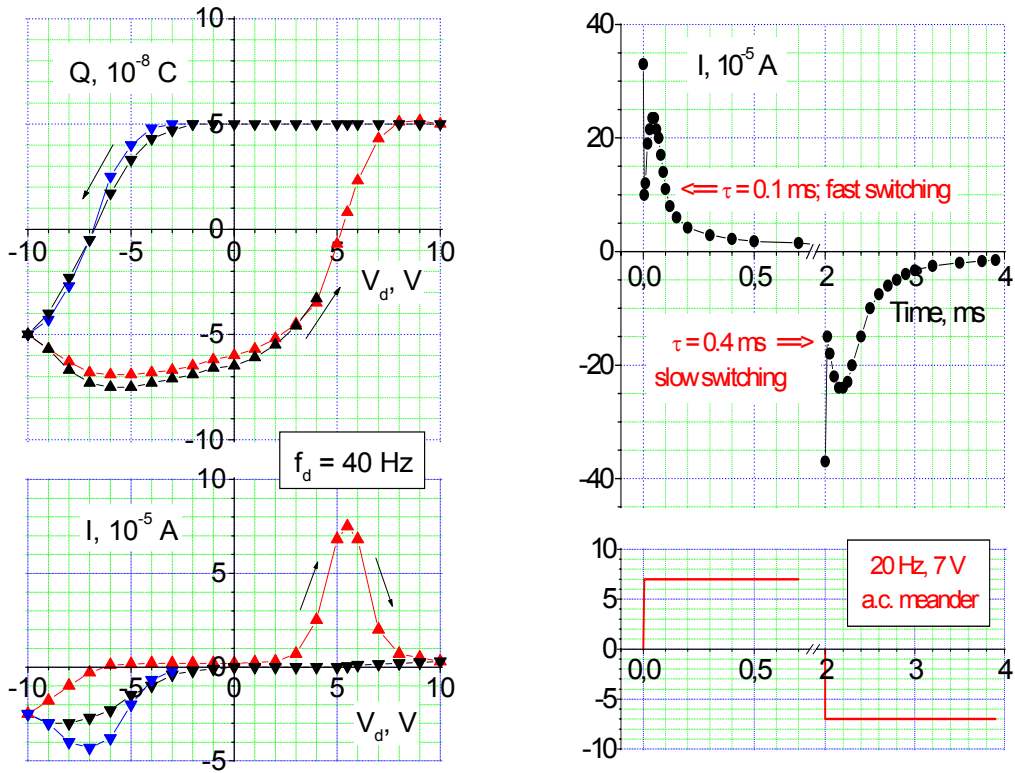

**Figure 6.** Polarization reversal and switching characteristics for one side water-ethanole damaged (negative branch) Ag-electroded TGS plate (0.3 mm thick):
 **(left)** charge-voltage (top) and current-voltage (bottom) loops;
 **(right)** transient currents (top) with schematic representation of drive voltage (bottom).

Comparison of Figs 4 and 5 for RPTC and Fig. 6 for TGS shows the correspondence of general peculiarities of I-V- and Q-V-loops and I-t-curves for RPTC and ferroelectric TGS samples.

Observed "humps" on the positive and negative branches of I-V-loops (Fig. 5) and the maxima on I-t-curves (Fig. 4) seems to be similar to the maxima of the currents of polarization reversal (Fig. 6, left) and currents of polarization switching (Fig. 6, right) of TGS samples characteristic for many ferroelectric materials.

The suppressed "humps" on I-V-loops (Fig. 5, left) and convex shape of Q-V-loops (Fig. 5, right) for RPTC are similar to those observed for the negative drive voltage runs of I-V-loops and Q-V-loops of TGS samples with one side water -ethanol damaged (Fig. 6, left).

It is necessary to note that the "humps" on bipolar I-V-loops (similar to shown in Figs 2 and 4) are characteristic for MIS-structures in the presence of mobile ionic charge (Chou 1971, Kuhn 1971). A general view of the unipolar I-V-curves (Fig. 3) is also similar to that observed for typical ionic conductor $RbAg_4I_5$ (Kukoz 1977). The existence of "humps" on I-t-curves is characteristic also for semiconductor systems under transfer of injected charge carriers (Lampert and Mark 1970).

## IV.2. On humidity sensing mechanism

The results obtained for the RPTC samples are more pronounced in comparison with those obtained for zeolite-like Al-Si-systems of Na-Y type (Bravina *et al.* 1999). These Al-Si-systems can be considered as relatively inert media for reactions connected with dissociation of water H-O-H molecules, electrochemical and ionic transport processes.

A protonic type of humidity sensing mechanism was recognised for porous oxides at room temperatures (Kulwicki 1991, Traversa 1995). It is also recognized that in case of humidity sensing ceramics with protonic conduction sensing mechanism, doping with alkali ions facilitates formation of hydrated protons. For example for $TiO_2$ based humidity sensors charge transport enhance was found when doping $TiO_2$ with 10 at % $Li^+$ or $K^+$ ions (Gusmano *et al.* 1996) or with $La^{3+}$ and $K^+$ ions (Anbia M. and Fard, S.E.M. 2011). It is also known (Kong *et al.* 1997) that, for example, similar to other porous ceramics with protonic type of humidity sensing mechanism, the humidity sensitivity of $SiO_2$ can be enhanced by adding electrolyte dopants, e. g., LiCl with alkali metal ions (Kong *et al.* 1997). The mechanism of charge transfer can be connected with the hopping transport by means of switching dangling bonds disposed on the surface as was shown for porous Si (Stievenard and Deresmes 1995).

For $Rb_4W_{11}O_{35}$ electronic and/or protonic/ionic transport processes with participation of $H^+$ and alkali metal $Rb^+$ cations of Rb sublattice can take place in addition to protonic transport due to dissociation of water molecules (H-O-H → $H^+$ + $OH^-$) in the water layer formed by physical adsorption.

Existence of half-occupied positions for interlayer $Rb^+$ ions in $Rb_4W_{11}O_{35}$ structure (Solodovnikov *et. al* 1998, Solodovnikov *et al.* 2002) (Fig. 1) supposes the possibility of its subsequent occupation, which can result in facilitation of protonic/ionic transfer with participation of these positions.

The similarity of peculiarities of I-t-curves (Fig. 4) and also I-V- and Q-V-loops (Fig. 5) for moistened RPTC and water damaged TGS (Fig. 6) evidences the ferroelectric-like reaction on the applied voltage. This conclusion is supported by the observed changes of the current maxima duration depending on humidity level in RPTC (see Fig. 4, right) and difference in times of polarization switching under different voltage polarity on water damaged and not damaged sides for TGS (see Fig. 6, right).

Such type of reaction seems to be connected with polarization phenomena as a result of disturbance of switching dangling bonds of surface conglomerate of RPTC by adsorbed dipolar $H_2O$ molecules (dipole moment 1.84 D ≈ 5.5·$10^{-30}$ C·m).

Water molecules adsorbed on the surface of RPTC change surface electrical conditions and seem to be a decisive factor in forming temporal polar media with increased dielectric permittivity, enhanced ionic transport and reversible polarization of ferroelectric-like or ionic space charge type.

## V.     Application aspect

Increase of humidity results in increase of bipolar I-V-loops slope and width and Q-V-loops size along Q-axis (see Fig. 5) and also I-t-curves swing.

Taking into consideration that for I-V-loop at $V_d = 0$ current value $I(V = 0) \sim C(V = 0) = C_0$ and the slope $dI(V_d)/dV_d \propto 1/R_0$, $R_0 = R(V = 0)$ it is possible to evaluate zero voltage capacitance $C_0$ and resistance $R_0$ values. So, from I-V-loops can be estimated relative changes of capacitance $\delta C_0 = C_0(H_r^{max})/C_0(H_r^{min})$ and resistance $\delta R_0 = R_0(H_r^{min})/R_0(H_r^{max})$ connected with $H_r$ rise from its minimal ($H_r^{min}$) to maximal ($H_r^{max}$) values. The values of $\delta C_0 \approx 14$ and $\delta R_0 \approx 10$ for $H_r^{max} = 95$ % and $H_r^{min} = 50$ % were estimated from the data in Fig. 4.

The values of $\delta C_0$ and $\delta R_0$ of order of 10 for the same $Rb_4W_{11}O_{35}$ sample were registered under sharp rise and subsequent decrease of $H_r$ value in the limits of (50 - 95) % by means of transformer RC-bridge (1 kHz) with oscilloscopic indication. The values of response time on wet air pulse and recovery time of restoring the initial C and G values are ~ 1 s and ~ 100 s respectively. So, $Rb_4W_{11}O_{35}$ ceramics can be considered as promising materials for creation of humidity sensitive elements for environment monitoring systems.

In particular, $C(H_r)$ and $R(H_r)$ changes of C and R caused by humidity $H_r$ variations could be used for controlling LC- or RC- frequency generator in the case of inserting the humidity sensitive element into corresponding frequency control circuits.

Variations of $C(H_r)$ were proposed for realizing the conversion of humidity changes into frequency ones on the base of radiofrequency resonance LC-bridge (600 kHz) with sound frequency output on the beats frequency between signals from reference and "humidity sensitive" circuits with various nano- and meso-porous systems based humidity sensors, including RPTC (Bravina *et al.* 1999, Bravina *et al.* 2000).

Direct humidity-to-frequency or humidity-to-time conversion was also realized by using $C(H_r)$ and $R(H_r)$ variations in RC-generator based on operational amplifier or on amplifier with dynamic load with "humidity sensitive" RC-elements in external feedback circuit.

## VI. Conclusion

For the investigated $Rb_4W_{11}O_{35}$ ceramic samples the peculiarities of I-V- and Q-V-loops and I-t-curves observed under high humidity impact are mainly connected with "products" of interaction of surface states of $Rb_4W_{11}O_{35}$ with aquatic adsorbate as a decisive factor in forming temporal polar protonic/ionic conducting media on the surface conglomerate – aquatic adsorbate boundary similar to 2D electrolytic bath.

$H_2O$ adsorption enhanced electronic and/or protonic/ionic space charge transfer with participation of $H^+$ and Rb ions (vacancies) and surface dangling bonds is supposed.

The similarity of peculiarities of I-V- and Q-V-loops and I-t-curves for RPTC and hydrogen-bounded ferroelectric TGS indicates the existence of ferroelectric-like type reaction on external voltage, which can be connected with polar character of surface dangling bonds disturbed by dipolar $H_2O$ molecules adsorbed on surface conglomerate of RPTC.

The set of data on transformations of I-V- and Q-V- loops, and I-t-curves of RPTC under pulse flux of wet air demonstrates the efficiency of using the methods of dynamic electrophysical characterization for studying impact of fast humidity changes on surface-active ceramic materials.

A reversible character and appropriate time scale of humidity impact induced changes of conductance and capacitance as well as the transformations of Q-V-loops, I-V- and I-t-curves of the selected RPTC samples are considered as promising for application of these materials in fast humidity sensing.

**Acknowledgments**

This work was partly supported by the Russian Foundation for Basic Research, Grant no. 03-08-00384.